\documentclass{CHEP2004}
\usepackage{graphicx}

\begin{document}

\title{The CEDAR Project}

\author{J. M. Butterworth, S. Butterworth and B. M. Waugh, UCL, London, UK\\
W.J. Stirling and M.R.Whalley\thanks{M.R.Whalley@durham.ac.uk}, IPPP, Durham, UK}

\maketitle

\begin{abstract}
We describe the plans and objectives of the CEDAR project (Combined e-Science Data 
Analysis Resource for High Energy Physics) newly funded by the PPARC e-Science programme
in the UK.
CEDAR will combine the strengths of the well established 
and widely used HEPDATA database of HEP data and the innovative JetWeb 
data/Monte Carlo comparison facility, built on the HZTOOL package, and will exploit 
developing grid technology. The current status and future plans of both of these 
individual sub-projects within the CEDAR framework are described, showing how they 
will cohesively provide (a) an extensive archive of Reaction Data, (b) validation and 
tuning of Monte Carlo programs against these reaction data sets, and (c) a validated 
code repository for a wide range of HEP code such as parton distribution functions 
and other calculation codes used by particle physicists. Once established it is 
envisaged CEDAR will become an important Grid tool used by LHC experimentalists in 
their analyses and may well serve as a model in other branches of science where there
is a need to compare data and complex simulations. 
\end{abstract}

\section{THE PHYSICS PROBLEM}

Particle physics experiments at high-energy accelerators provide a wealth of data on
the final state in electron-positron, lepton-proton and proton-(anti)proton interactions.
These data represent a triumph for the Standard Model, particularly in precision
electroweak measurements and the verification of the QCD sector to an impressive degree 
of precision.

\begin{figure}[htb]
\centering
\includegraphics*[width=65mm]{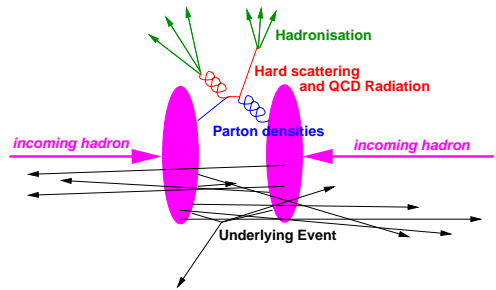}
\caption{The complexity of a collision process.}
\label{complexity}
\end{figure}

Despite these successes, several aspects of high-energy collisions are still poorly 
understood, often due to technical difficulties in the calculation of non-perturbative
or complex perturbative effects. Such lack of understanding can be a limiting
factor in the accuracy of new measurements.  Examples with particular relevance to the
LHC include parton distribution functions (PDFs) in hadrons, hadronisation in the
final state, multijet production and \lq\lq underlying events". Fig.~\ref{complexity} illustrates 
some of these processes in a complex high energy event. All these processes are 
calculated and/or modelled in Monte Carlo simulation and calculation programs that
employ fits to existing data.  Consistent tuning of the free parameters of these models,
and confirmation of the physics assumptions they contain, is a non-trivial matter
since the measurements are made with a variety of colliding beams, in many different
regions of phase space, and for many complex observables. 
 
\section{THE SOLUTION -- CEDAR}

A solution to the above problem will be provided by CEDAR, a project newly funded 
by the PPARC e-Science programme in the UK, 
which will construct a
resource for particle physics enabling the predictions of
Monte Carlos, and other calculation programs, to be easily 
compared with real data.  CEDAR will allow the parameters of the
models to be varied and to be simultaneously compared to as wide a range of data distributions
as is necessary to maintain global consistency. These global comparisons are vitally important 
as it is quite possible to obtain a good fit to a new set of data but at the same time lose 
the quality of fit to other existing data.

CEDAR  will combine the strengths of the established and widely used HEPDATA database \cite{hepdata} 
of high-energy physics data
and the innovative JetWeb data comparison facility \cite{jetweb2}, and will exploit 
developing Grid technology. 

In short:- 
$$
\mathit{CEDAR} = \mathit{JetWeb} + \mathit{HEPDATA}  + \mathit{more...}
$$
The individual parts, including the \lq\lq{\it ...more}" are described below.

\subsection{JetWeb}

\begin{figure*}[t]
\centering
\includegraphics*[width=125mm]{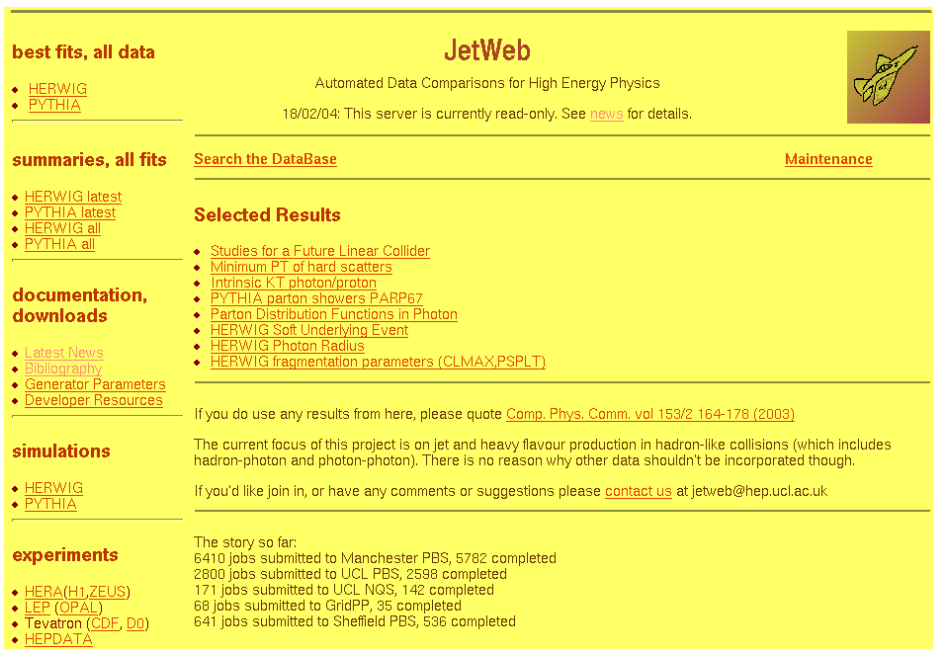}
\caption{The JetWeb home web page}
\label{jetwebhome}
\end{figure*}
 
JetWeb is a WWW interface, developed at UCL, which provides a facility for direct
comparison between the predictions of Monte Carlo programs and measured physics
distributions from experiments.  It is based upon the HZTOOL program \cite{hztool}
which generates the physics distributions from a given Monte Carlo program
by calculating the relevant variables and applying the cuts used in the published measurements.

The JetWeb interface allows the user to request and display quantitative 
comparisons of chosen Monte Carlo models with any number of measured data 
distributions.  In this 
way the compatibility with the older data sets can be maintained when tuning to new
data sets. 

Since the generation of Monte Carlo events to calculate the predicted distributions to the required 
accuracy is a very compute intensive operation, 
JetWeb maintains a database (MySQL) of Monte Carlo generated data which can queried, 
or added to at user request, thus increasing the available statistics.
JetWeb is already \lq\lq grid-enabled" and has submitted its Monte Carlo jobs to GridPP 
with successful outcomes, as well as directly using the computer systems at UCL, 
Manchester and Sheffield.  

Fig.~\ref{jetwebhome} shows the JetWeb WWW home page which provides the user with
the starting point to begin a new fit, or to search the JetWeb database for results from
previous fits.

\begin{figure}[htb]
\centering
\includegraphics*[width=65mm]{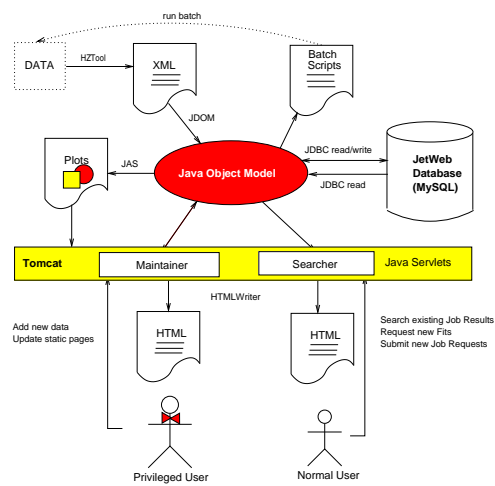}
\caption{The various components of JetWeb }
\label{jetweb}
\end{figure}

\begin{figure*}[t]
\centering
\includegraphics*[width=125mm]{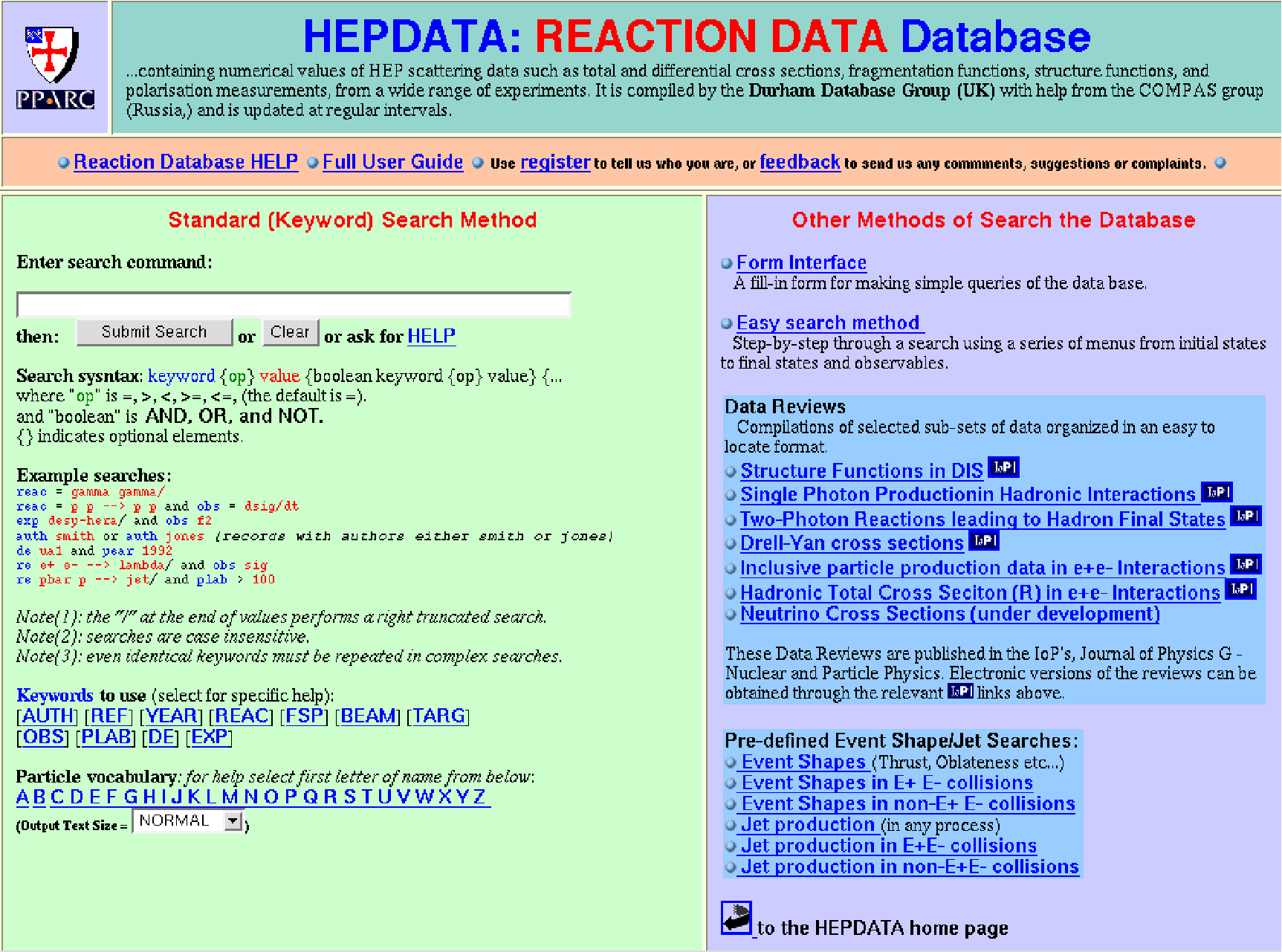}
\caption{The HEPDATA Reaction Data page }
\label{hepdatahome}
\end{figure*}

Fig.~\ref{jetweb} show the main components of the JetWeb server. Interaction with
the user is via two Java Servlets, running in a Tomcat servlet container linked
to an Apache web server: the Searcher servlet provides the the standard functions
 available to a general user;
the Maintainer provides additional functions, such as adding new data, to the
JetWeb administrators.

At the heart of JetWeb is the Java Object Model (JOM) which contains the properties and interactions of the
Model, Papers, Plots and Fits.  (In this context a Model means a unique generator version and set of
parameters.) Data is converted between formats via this model, e.g. from Database to HTML display.
The JOM interacts with the JetWeb database, a MySQL database which  stores 
experimental data as well as predicitions from various models.
The predictions are generated by running the Monte Carlo generators within the HZTOOL package,
and the results are available to the user who requested the run and to any future user who makes
a similar request.

At present, the size of the JetWeb database is restricted by the relatively small proportion
of experimental data distributions which have HZTOOL routines to generate the predicted distributions from 
Monte Carlos generators.

CEDAR will  do the following for JetWeb:-
\begin{Itemize}
\item  Re-design HZTOOL in OO structured C++ for long term maintenance and development.
This is vitally important as new HEP code, and in particular new Monte Carlos, are being written using 
C++ in such ways.  
\item  Incorporate new Monte Carlos.  At present only HERWIG and PYTHIA are available.
\item  Use HEPDATA as the source of \lq\lq real" data distributions, giving
access to more data including (eventually) LHC data.  
\item  Develop the Web and Grid interface to the model validator data.

\end{Itemize}
It is stressed however that although direct access to the data from HEPDATA will increase the
scope of experimental data available to JetWeb, it will always be necessary for any data to be
used to have a corresponding HZTOOL (or some equivalent) routine.  The CEDAR project will 
actively encourage the participation of the experimental community in producing these routines.
  
\subsection{HEPDATA}

HEPDATA is a PPARC funded project which has been in existence now for over
25 years.  Its principal aim has remained essentially the same over this period,
namely to compile scattering data from all types of HEP reactions (cross sections,
event shapes, polarisations, etc...) and to make the resulting compilations easily available to 
the whole community.

More recently other services such as the hosting of mirrors of the SLAC SPIRES
databases and the Berkeley Particle Data Group (PDG)
Review of Particle Physics (RPP) web pages in the UK,
have been added to the HEPDATA operation. 
HEPDATA also provides a unique and comprehensive 
PDF code server with an on-line PDF calculation, display and comparison facility 
These are all accessible from the main HEPDATA home 
web page, shown in Fig.~\ref{hepdatahome}.

The scope of the HEPDATA database covers cross sections from all types of particle 
physics reactions. 
It is emphasised that it does not contain \lq\lq particle properties" which fall
into the domain of the RPP of the Berkeley PDG. It also not contain raw data such as
found on DSTs of experiments.  To appear in the database the data are generally in the final published
form. Ideally, to be most useful,  they should be fully corrected for acceptances and efficiencies and 
be model independent.
The database contains data from around 10000 publications dating from early experiments in
the 1970s to the present day data from the LEP, Tevatron and HERA collaborations.
It is regularly added to and updated.  
The data are obtained
from journals and preprints and direct from the experiments 
especially when data appear only in graphical form in a publication. In the latter case the authors of the
paper are contacted to obtain the exact numerical values shown in the plot. It is
very important that this is done at the time of the publication as experience has
shown how difficult it is to obtain numerical values at a later date. Data are
rarely read from plots due to the difficulty in getting accurate representations
Finally, verification of 
data entered into the database is always sought from the experimenters 
themselves.

At present the HEPDATA project uses a non proprietary database management system
(BDMS - the Berkeley Database Management System), as it has since its inception.
This is a hierarchical DBMS in which data are stored in a tree like structure with 
the paper as the main record unit and the all data tables in a particular publication
stored within that one unit (a data record).  
While this DBMS has proved very resilient and stable over the
years, it is clearly not suitable for the present purpose of directly interfacing with 
JetWeb, or with any other resource over the Internet/Grid.

CEDAR will  do the following for HEPDATA:-
\begin{Itemize}
\item Migrate the data from the HEPDATA BDMS hierarchical database 
to a MySQL relational database.  Not only will this provide the
necessary means to interface with JetWeb,and other resources if necessary, 
but it will also address the long term future needs of HEPDATA by using 
a DBMS which is more standard and maintainable than its present one.
  
\item Integrate the new database into JetWeb as the
source of its \lq\lq real" data.   This will expand the number of 
available data sets to which JetWeb has access.

\item Make the new database available on the Internet/Grid as a 
networked database for general use. This will involve implementing 
and expanding on
the existing functionality of the HEPDATA web search and display methods. Expansions
envisaged include direct access to the data through the conventional networks and
also via the \lq\lq Grid".

\item Develop new methods of direct entry and validation of
data by the experiments, thus making them in control of their own data.
At present the entry and verification of data in the database is generally instigated
and controlled by the HEPDATA personnel.  In future CEDAR will seek methods and
formats (e.g. XML) in which experiments can enter and maintain their own data in
the database potentially using Grid technologies and access validation methods.
It should also be noted that the development of an XML document format for transfer
of particle physics data will be useful also for data output to the Grid etc. 
as well as input to HEPDATA.
\end{Itemize}

\subsection{more... HEPCODE}

As well as the improvements to JetWeb and HEPDATA and their integration
as discussed in the previous sections, it is also the intention of CEDAR
to provide access to current and validated versions of various HEP
theory and experiment software used at present, and in the future LHC era.
A partial list of such software includes JETRAD, DYRAD, EXCALIBUR, (DI)PHOX,
ZFITTER, RACOONWW and MADGRAPH.  No central repository of such codes 
currently exists in an organised and consistent way at present and 
as such it will be of great benefit to 
the community.    

A prototype version, HEPCODE, has been set up on the IPPP website (http://www.ippp.dur.ac.uk/hepcode/)
at Durham which includes, as well as a table of available programs and links, a web
form for submission of new codes to the repository.
\section{SUMMARY}

In summary we list the various tasks the CEDAR project will be engaged upon:-
\begin{Itemize}

\item Convert HEPDATA to a relational MySQL database.

\item Modify JetWeb to access data direct from HEPDATA.

\item Design and implement an OO replacement for HZTOOL.

\item Extend the number of HZTOOL routines/models available in JetWeb.

\item Include new generation C++ Monte Carlos.

\item Develop CEDAR grid tools for automatic validation and data access.

\item Develop standard document formats (e.g. XML) for experimental
physics data and results, which can be used for example as methods for reaction data 
export/import from experiments, as well as being useful in a more general context.

\item Develop a code repository for a wide range of HEP codes and integrate with the
validation centre.

\item Incorporate LHC and, eventually, ILC data.
\end{Itemize}

Once established and running with all the latest codes and data sets, it is envisaged that CEDAR will
become an important tool used by LHC experimentalists in their analyses. Other branches of science
which have need to compare data and complex model situations may also find the techniques and tools
used in CEDAR of benefit to their work.

\section{THE CEDAR TEAM}

\noindent {\bf At UCL:-}

Jon Butterworth - J.Butterworth@ucl.ac.uk

Susanna Butterworth - sb@hep.ucl.ac.uk

Ben Waugh - waugh@hep.ucl.ac.uk

\noindent {\bf At Durham:-}

James Stirling - W.J.Stirling@durham.ac.uk

Mike Whalley - M.R.Whalley@durham.ac.uk

A N Other - to be appointed from 1st April 2005

\noindent {\bf Web Links:-}

CEDAR - http://cedar.ac.uk/

JetWeb - http://jetweb.hep.ucl.ac.uk/

HEPDATA - http://durpdg.dur.ac.uk/hepdata/

HZTOOL - http://hztool.hep.ucl.ac.uk/
\section{ACKNOWLEDGEMENTS}

The CEDAR project is funded by the PPARC e-Science grant PP/B5027/44/1.
HEPDATA is funded by PPARC grant PP/B500590/1.

\end{document}